\documentclass[prl,twocolumn,showpacs,preprintnumbers,floatfix,amsmath,amssymb]{revtex4}

\usepackage{graphicx}% Include figure files
\usepackage{dcolumn}% Align table columns on decimal point
\usepackage{bm}% bold math

\newcommand{\Rend}{\ensuremath{R}}
\newcommand{\Rgyr}{\ensuremath{R_\mathrm{g}}}

\newcommand{\dpm}{\ensuremath{d_\mathrm{p}}}
\renewcommand{\vec}[1]{\ensuremath{\mathbf{#1}}}

\begin{document}

\title{Perimeter Length and Form Factor of Two-Dimensional Polymer Melts}

\author{H. Meyer, T. Kreer, M. Aichele, A. Cavallo, A. Johner, J. Baschnagel, and J.P. Wittmer}

\affiliation{Institut Charles Sadron, 23 rue du Loess, BP 84047, 67034 Strasbourg Cedex 2, France}

\date{\today}

\begin{abstract} 
Self-avoiding polymers in two-dimensional ($d=2$) melts are known to adopt compact 
configurations of typical size $R(N) \sim N^{1/d}$ with $N$ being the chain length. 
Using molecular dynamics simulations we show that the irregular shapes of these chains
are characterized by a perimeter length $L(N) \sim R(N)^{\dpm}$ of fractal dimension 
$\dpm = d-\Theta_2 =5/4$ with $\Theta_2=3/4$ being a well-known contact exponent.
Due to the self-similar structure of the chains, compactness and perimeter fractality repeat 
for subchains of all arc-lengths $s$ down to a few monomers. The Kratky representation of the 
intramolecular form factor $F(q)$ reveals a strong non-monotonous behavior with 
$q^2F(q) \sim 1/(qN^{1/d})^{\Theta_2}$ in the intermediate regime of the wavevector $q$.
Measuring the scattering of labeled subchains %($s F(q) \sim L(s)$) 
the form factor may allow to test our predictions in real experiments. 
\end{abstract}


\maketitle

\begin{figure}[tb]
\includegraphics*[width=8.0cm]{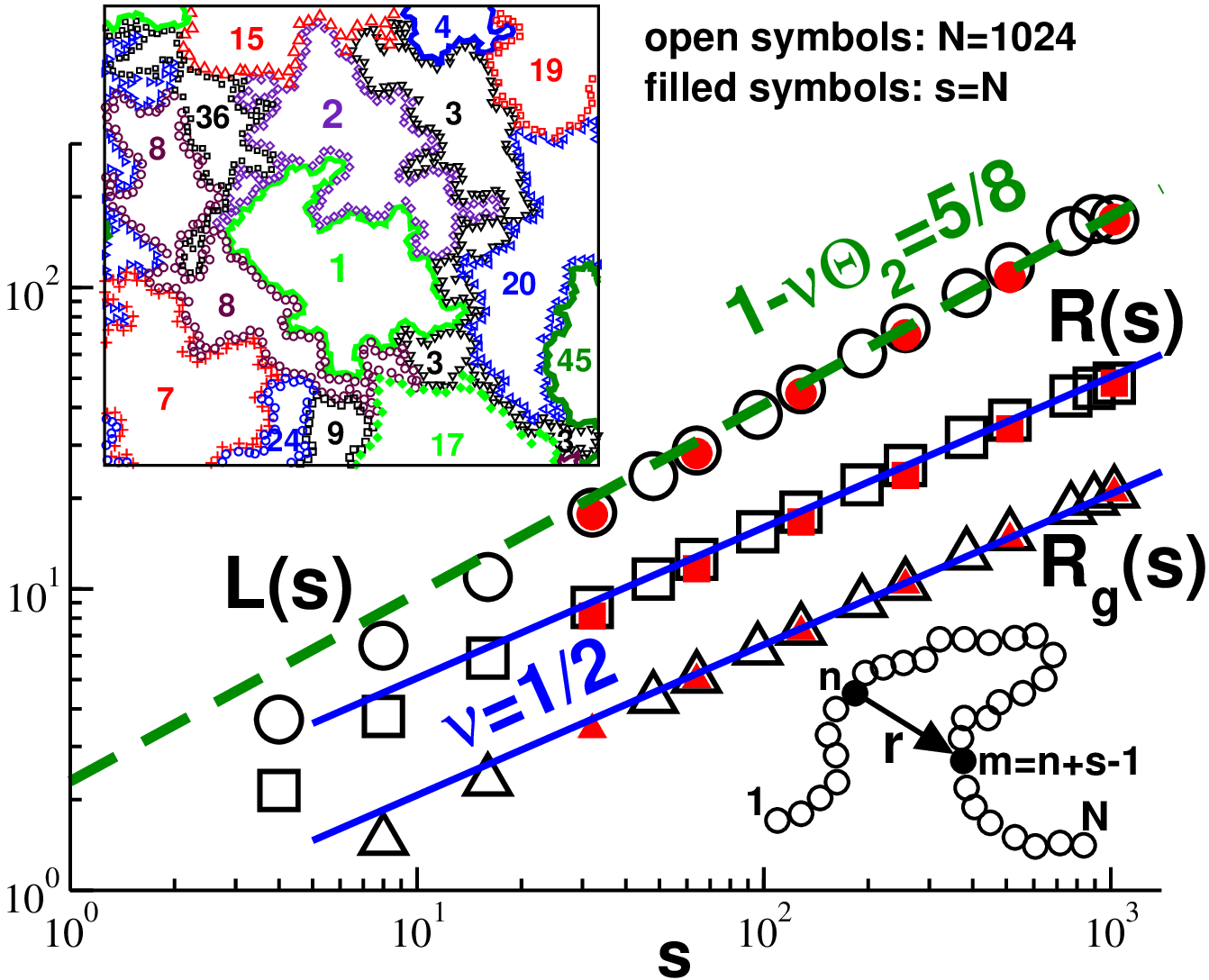}
\caption{(color online).
A snapshot of a melt configuration with chain length $N=1024$ and monomer density $\rho=7/8$
can be seen in the left panel. Only the perimeter monomers interacting with other chains are indicated.
The numbers refer to an arbitrary chain index used for computational purposes.
The chains are compact, i.e., they fill space densely, but compactness does not imply
a disk-like shape (see, e.g., chains 3 and 8). 
The main figure presents 
the end-to-end distance $R(s)  = \langle \vec{r}^2 \rangle^{1/2}$,
the radius of gyration $\Rgyr(s)$, 
and the mean perimeter length $L(s)$ of segments 
containing $s=m-n+1$ monomers as indicated by the sketch on the right.
The full symbols refer to overall chain properties ($s=N$).
The solid lines confirm the exponent $\nu=1/2$ for the segment size,
the dashed line confirms the scaling of $L(s)$ suggested by Eq.~(\ref{eq_fractal}).
\label{fig_sum}
}
\end{figure}

%
%%%%%%%%%%%%%%%%%%%%%%%%%%%%%%%%%%%%%%%%%%%%%%%%%%%%%%%%%%%%%%%%%%%%%%%%%%%%%%%%%%%%%%%%%%%%%%%%%%%%%%%
{\em Introduction.}
It is well known that linear polymers in two dimensions (2d) adopt compact and segregated 
conformations at high densities \cite{DegennesBook,dupl,semenov,carm,cav,yethiraj}. 
This is expected to apply not only on the scale of the total chain of $N$ monomers but also 
to subchains comprising $s$ monomers, at least as long as the segments are not too small 
($1 \ll s \le N$). The typical size $R(s)$ of a chain segment should thus scale as 
$R(s) \sim s^{\nu}$ with 
%a Flory exponent $\nu=1/d$ 
an exponent $\nu=1/d$
set by the spatial dimension $d=2$.  
Compactness does obviously not imply Gaussian chain statistics \cite{dupl,semenov} nor does segregation 
of chains and chain segments impose disk-like shapes minimizing the average perimeter length $L(s)$ of 
chain segments. The boundaries of chains and of chain segments are in fact found to be highly irregular 
as revealed by the snapshot presented in Fig.~\ref{fig_sum}. 
Using scaling arguments and molecular dynamics simulations we show below that these 
perimeters are {\em fractal}, scaling as
\begin{equation}
L(s) \sim R(s)^{\dpm} \sim s^{1- \nu \Theta_2}
\label{eq_fractal}
\end{equation}
with $\dpm = d- \Theta_2 = 5/4 > 1$ being the fractal line dimension. Our work is based on the pioneering 
work by Duplantier who predicted a contact exponent $\Theta_2 = 3/4$ \cite{dupl} characterizing the 
intrachain segmental size distribution. 
In contrast to many other possibilities to characterize numerically the compact chain 
conformations the perimeter length %is of interest since it 
can be related to the intrachain form factor $F(q)$ making it accessible experimentally, 
at least in principle, by means of small-angle scattering experiments 
\cite{BenoitBook,mai99}.

%
%%%%%%%%%%%%%%%%%%%%%%%%%%%%%%%%%%%%%%%%%%%%%%%%%%%%%%%%%%%%%%%%%%%%%%%%%%%%%%%%%%%%%%%%%%
%{\em Outline.}
We recall first the computational model used for this study and confirm then numerically the scaling of 
$R(s)$ and $L(s)$ suggested above. 
%The scaling analysis of various intrachain 
%properties such as the segmental size distribution $G(r,s)$, 
%the bond-bond correlation function $P_2(s)$, and the intrachain 
%form factor $F(q)$ will allow us to derive Eq.~(\ref{eq_fractal}).
%We finish the paper by discussing possible consequences for the dynamics of 2d melts.
The analysis of intrachain 
properties such as the segmental size distribution, 
the bond-bond correlations, and the intrachain 
form factor will allow us to demonstrate Eq.~(\ref{eq_fractal}).
We conclude by discussing consequences for the dynamics of 2d melts.

%
%%%%%%%%%%%%%%%%%%%%%%%%%%%%%%%%%%%%%%%%%%%%%%%%%%%%%%%%%%%%%%%%%%%%%%%%%%%%%%%%%%%%%%%%%%

%%%%%%%%%%%%%%%%%%%%%%%%%%%%%%%%%%%%%%%%%%%%%%%%%%%%%%%%%%%%%%%%%%%%%%%%%%%%%%%%%%%%%%%%%%%%%%%%%%%%%%%%%%%%%%%
{\em Computational issues.}
Our numerical results are obtained by molecular dynamics (MD) simulations
of monodisperse linear chains at high densities. 
%We solve the classical equations of motion of the multichain 
%system via the Velocity-Verlet algorithm at constant temperature using a Langevin thermostat 
%\cite{FrenkelSmitBook}.
%
Our coarse-grained polymer model Hamiltonian is essentially identical to the well established 
Kremer-Grest (KG) bead-spring model \cite{KG86,KG90,LAMMPS} where the excluded volume 
interaction among monomers is mimicked by a purely repulsive Lennard-Jones potential and the 
chain connectivity is assured by harmonic springs calibrated to the 
``finite extendible nonlinear elastic" (FENE) springs of the KG model.
We focus in this work on melts of density $\rho=7/8$ at temperature $k_BT=1$ with chain lengths
ranging up to $N=2048$ using a periodic simulation box of linear length $335.2$ containing $98304$ 
monomers \cite{foot_para}.

%
%%%%%%%%%%%%%%%%%%%%%%%%%%%%%%%%%%%%%%%%%%%%%%%%%%%%%%%%%%%%%%%%%%%%%%%%%%%%%%%%%%%%%%%%%%%%%%%%%%%%%%%%%%%%%%%
{\em Mean segment size and perimeter length.}
The main part of Fig.~\ref{fig_sum} presents the typical size and perimeter length of a chain segment 
between the monomers $n$ and $m=n+s-1$ as indicated by the sketch. 
Following \cite{WMBJOMMS04,WBM07} we average over all pairs $(n,m)$ possible in a chain of length $N$.
Averaging only over segments at the curvilinear chain center slightly reduces chain end effects,
however the difference is negligible for the larger chains, $N > 256$, we will focus on.
Open symbols refer to segments of length $s \le N$ of chains of length $N=1024$, 
full symbols to total chain properties ($s=N$).
The segment size may by characterized by either the second moment $R^2(s) = \langle \vec{r}^2 \rangle$ 
%of the distribution $G(\vec{r},s)$ 
of the end-to-end vector $\vec{r}$ of the segment (squares) or 
by its radius of gyration $\Rgyr^2(s)$ (triangles) \cite{DoiEdwardsBook}.
In agreement with various numerical studies \cite{carm,cav,yethiraj} the presented data confirms
that the chains are compact, i.e. $\nu=1/2$ (solid lines), on all scales $s$
\cite{foot_corrections}.
A perimeter monomer of a chain segment is defined as a monomer being within a distance $1.2$ to 
a monomer {\em not} belonging to the same chain segment \cite{foot_Ldistance}.  The mean number $L(s)$ of 
these perimeter monomers increases with a power-law exponent $1-\nu\Theta_2=5/8$ (dashed line) 
which is in perfect agreement with Eq.~(\ref{eq_fractal}), and this holds again on all scales for 
arbitrary segment lengths provided that the segment is sufficiently large ($s > 50$). 
We demonstrate in the following where the suggested scaling stems from.

\begin{figure}[tb]
\includegraphics*[width=8.0cm]{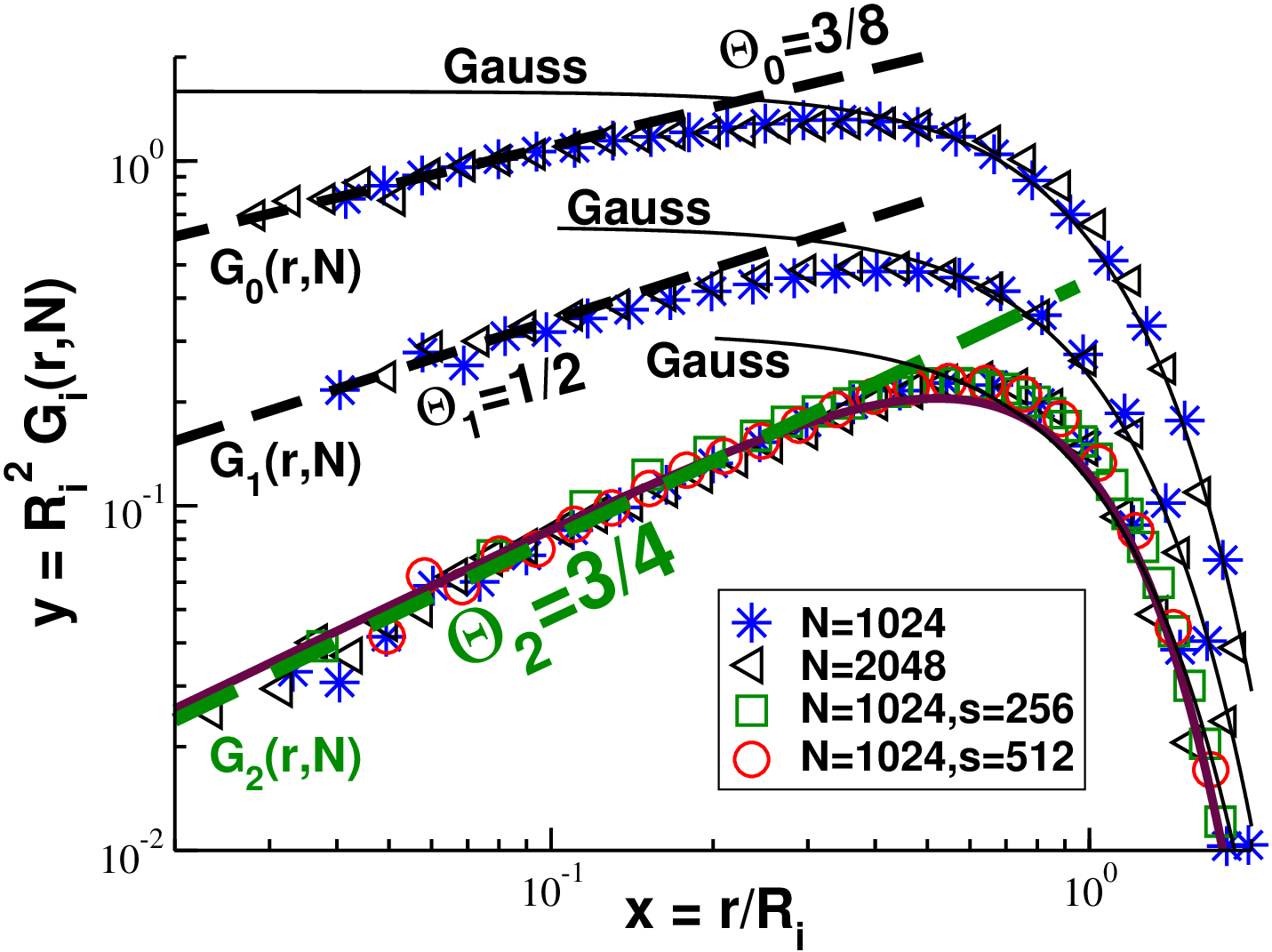}
\caption{(color online).
Scaling plot of various distributions of the end-to-end vector $\vec{r} = \vec{r}_m - \vec{r}_n$ 
of chain segments of chains of length $N=1024$ (stars) and $N=2048$ (triangle):
$G_0(r,N)$ for $n=1$ and $m=N$,
$G_1(r,N)$ for $n=1$ and $m=N/2$,
$G_2(r,N)$ for $n=N/4$ and $m=3N/4$.
$G_0(r,N)$ and $G_1(r,N)$ are shifted vertically for clarity.
The segmental size distribution $G(r,s)$ averaging over all pairs $(n,m)$
given for $N=1024$ with $s=256$ (squares) and $s=512$ (spheres) scales as $G_2(r,N)$.
The thin lines indicate the Gaussian distribution $y = \exp(-x^2)/\pi$ expected for ideal chains in 2d.
The power laws $y \approx x^{\Theta_i}$ (dashed lines)
observed for $x \ll 1$ have been predicted by Duplantier \cite{dupl}.
The Redner-des Cloizeaux formula for $G_2(r,N)$ is indicated by a solid line.
\label{fig_theta}
}
\end{figure}

{\em Segment size distributions and contact exponents.}
Obviously, the mere fact that 
the exponent $\nu$ 
%the Flory exponent $\nu$ 
is the same in 2d and 3d does not imply
that 2d melts are Gaussian \cite{dupl}.  This can be directly seen from the different probability
distributions of chain segment vectors $\vec{r} = \vec{r}_m - \vec{r}_n$ presented in Fig.~\ref{fig_theta}.
To simplify the plot we focus on the two longest chains simulated, $N=1024$ and $N=2048$.
$G_0(r,N)$ characterizes the distribution of the total chain end-to-end vector ($n=1$, $m=N$), 
$G_1(r,N)$ the distance between a chain end and the monomer in the middle of the chain ($n=1$, $m=N/2$) 
and $G_2(r,N)$ the distribution of an inner segment vector between the monomers $n=N/4$ and $m=3N/4$.
In addition, we indicate the segmental size distribution $G(r,s)$ averaging over all pairs
$(n,m)$ which has been used recently to characterize deviations from ideal chain behavior 
in 3d melts \cite{WBM07}.
All data for different $N$ and $s$ collapse on three distinct master curves if the axes are 
made dimensionless using the second moment $R_i^2$ of the respective distribution as indicated in the
figure. The only relevant length scale is thus the typical size of the segment itself. 
The distributions are not monotonous and are thus qualitatively different from the Gaussian
(thin lines) expected for random walks.
In agreement with Duplantier \cite{dupl} we find
\begin{equation}
R_i^d G_i(r,N) = x^{\Theta_i} f_i(x)
\label{eq_theta}
\end{equation}
with $x=r/R_i$ being the scaling variable and the contact exponents $\Theta_0=3/8$,
$\Theta_1=1/2$ and $\Theta_2=3/4$ (dashed lines) describing the small-$x$ limit where 
the universal functions $f_i(x)$ become constant. 
Especially the largest of these exponents, $\Theta_2$, is clearly visible. The contact 
probability for two monomers of a chain in a 2d melt is 
thus strongly suppressed compared to ideal chain statistics ($\Theta_0=\Theta_1=\Theta_2=0$).
As can be seen, the rescaled distributions $G(r,s)$ and $G_2(r,N)$ become identical 
for intermediate segment length, $1 \ll s \ll N$.
(Obviously, $G(r,s) \approx G_0(r,N)$ for very large segments $s \rightarrow N$.)
It is for this reason that the exponent $\Theta_2$ is the most important one for asymptotically
long chains where chain end effects can be neglected.
The rescaled distributions show exponential cut-offs for large distances. 
The Redner-des Cloizeaux formula \cite{DescloizBook} is a useful interpolation
formula which supposes that $f_i(x) = C_i \exp(-K_i x^2)$ with constants $C_i$ and 
$K_i=1+\Theta_i/2$ imposed
by the normalization and the second moment of the distributions \cite{ever}. This formula is by 
no means rigorous but yields reasonable parameter free fits as shown for $f_2(x)$.

%%%%%%%%%%%%%%%%%%%%%%%%%%%%%%%%%%%%%%%%%%%%%%%%%%%%%%%%%%%%%%%%%%%%%%%%%%%%%%%%%%%%%%%%%%%%%%%%%%%%%%%%%%%%%%%
\begin{figure}[tb]
\includegraphics*[width=8.0cm]{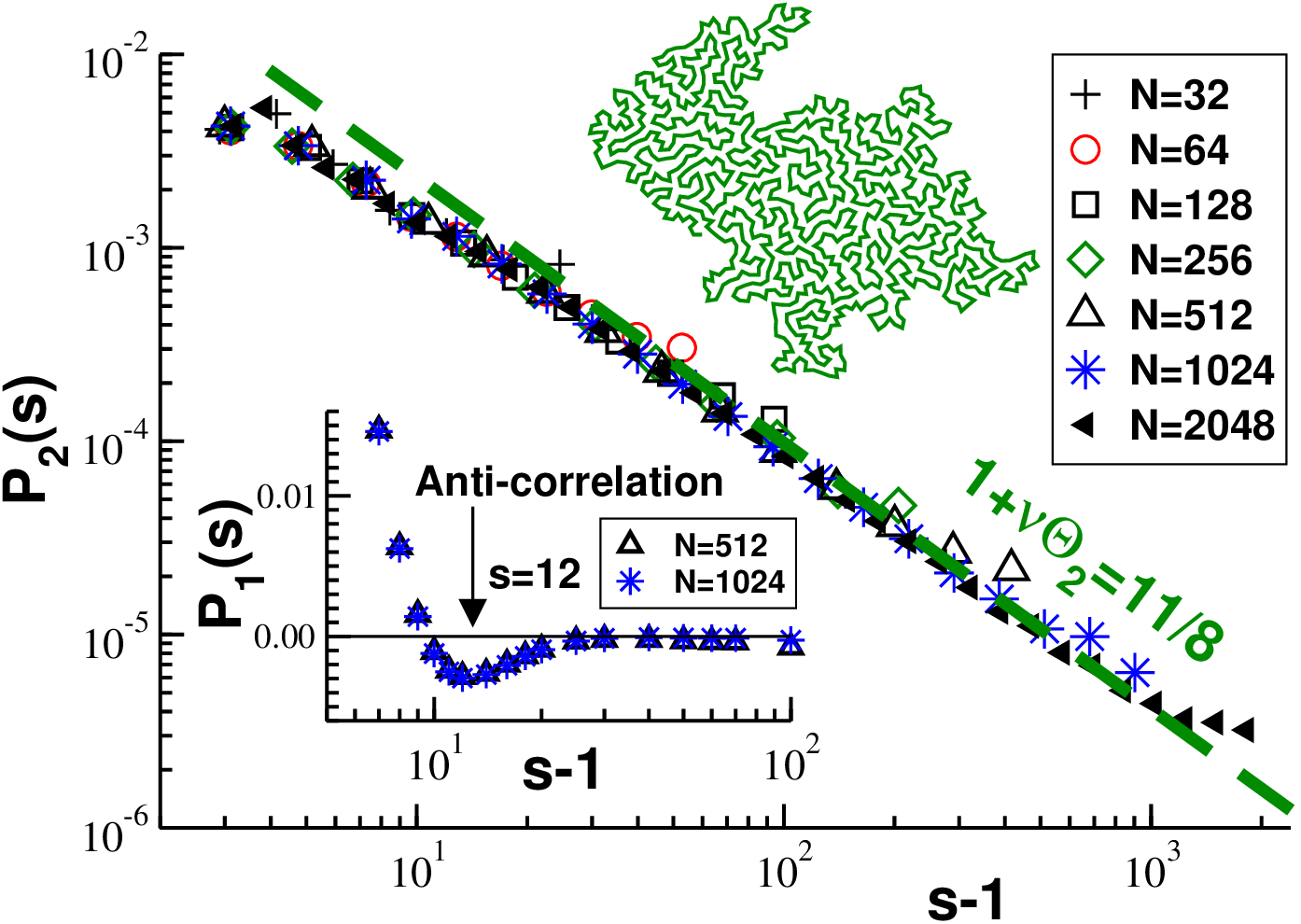}
\caption{(color online).
\label{fig_eiej} 
Bond-bond correlation functions $P_1(s) = \langle \vec{e}_n \cdot \vec{e}_m \rangle$ and 
$P_2(s) = \langle \left(\vec{e}_n \cdot \vec{e}_m\right)^2 \rangle -1/2$ {\em vs.} curvilinear
distance $s-1$ between normalized bond vectors $\vec{e}_n$ and $\vec{e}_m$. 
The first Legendre polynomial $P_1(s)$ (inset) shows an anti-correlation at $s \approx 12$.
The second Legendre polynomial $P_2(s)$ decays over two orders in magnitude as a power law (dashed line)
with an exponent $1+\nu\Theta_2=11/8$ in agreement with the return probability calculated from 
Eq.~(\ref{eq_theta}). 
}
\end{figure}

{\em Bond-bond correlations.}
The bond-bond correlation function $P_1(s) = \langle \vec{e}_n \cdot \vec{e}_m \rangle$ 
($\vec{e}_i$ denoting the normalized bond vector connecting the monomers $i$ and $i+1$)
has been shown to be of particular interest for characterizing the deviations from 
random walk statistics in 3d polymer melts \cite{WMBJOMMS04,WBM07}. The reason for this is that
$P_1(s)$ is proportional to the second derivative of the segment size $R(s)^2$ with
respect to segment length $s$ so that small deviations from the asymptotic
exponent $2\nu=1$ are emphasized \cite{WBM07}. 
As can be seen from the inset in Fig.~\ref{fig_eiej} deviations of this kind are small 
for large $s$ and may be neglected for the present study. The main effect visible is an 
anticorrelation at $s \approx 12$ due to the backfolding of the chain contour which can 
be directly seen from the snapshot.
Conceptually more important is the fact that the second Legendre polynomial 
$P_2(s) = \langle \left(\vec{e}_n \cdot \vec{e}_m\right)^2 \rangle -1/2$
reveals a clear power law behavior over two orders of magnitude in $s$ 
(dashed line).
The power law is due to the alignment of two bonds if they are sufficiently close,
i.e. the exponent measures the return probability after $s$ steps.
It follows from Eq.~(\ref{eq_theta}) that for $1 \ll s \ll N$ this is given by
$\lim_{r\rightarrow 0} G(r,s) \sim 1/s^{1+\nu\Theta_2}=s^{-11/8}$.
The agreement of the data with this exponent is excellent and provides
an independent confirmation of $\Theta_2=3/4$.

%%%%%%%%%%%%%%%%%%%%%%%%%%%%%%%%%%%%%%%%%%%%%%%%%%%%%%%%%%%%%%%%%%%%%%%%%%%%%%%%%%%%%%%%%%%%%%%%%%%%%%%%%%%%%%%
\begin{figure}[tb]
\includegraphics*[width=8.0cm]{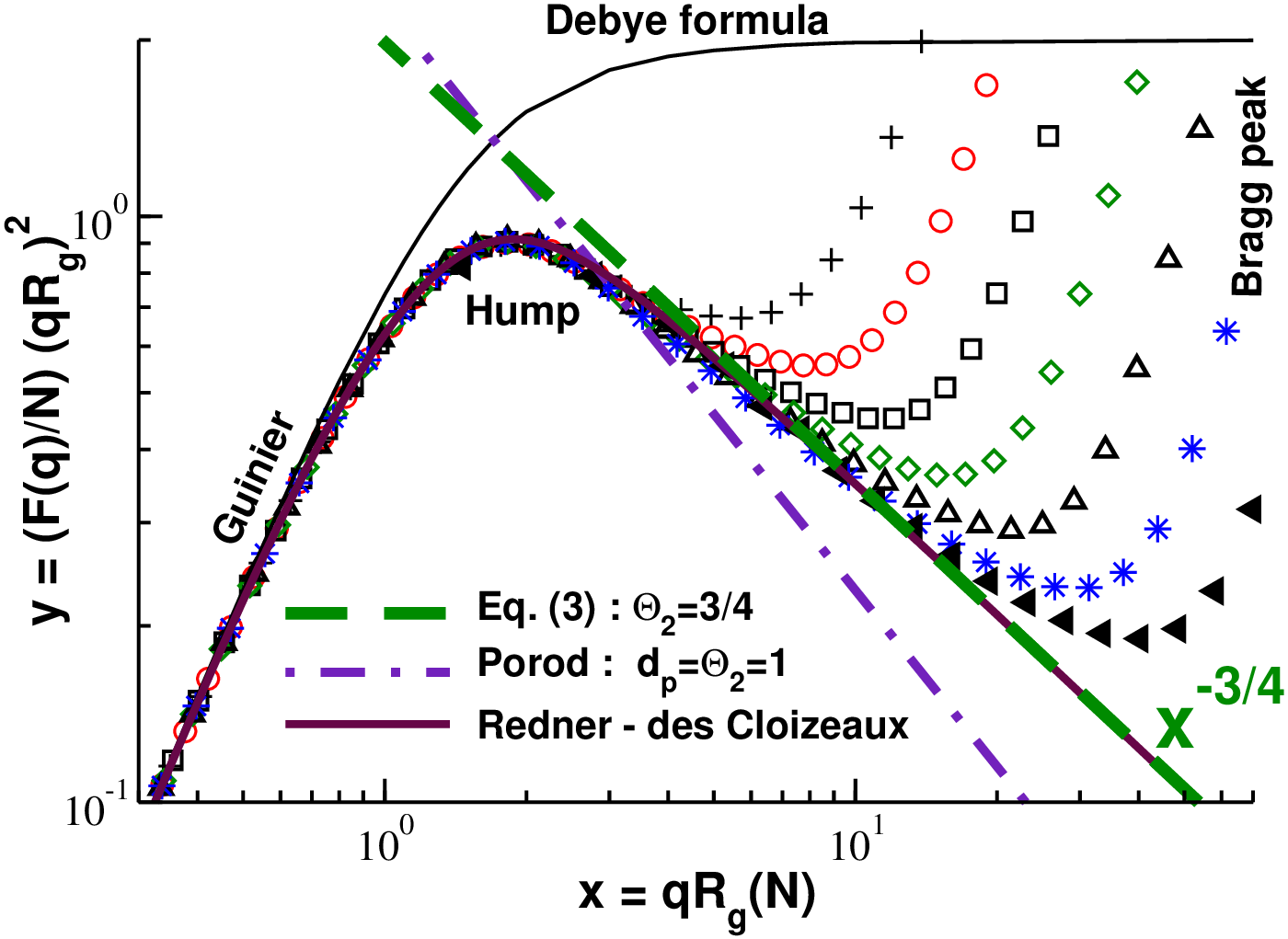}
\caption{(color online).
Kratky representation of the intramolecular form factor $F(q)$ as a function of $x=q\Rgyr(N)$
for different chain length $N$ using the same symbols as in Fig.~\ref{fig_eiej}. 
The Debye formula (thin line) corresponds to a chain length independent plateau for $x \gg 1$. 
By contrast to this, a strong non-monotonous behavior is revealed by our data
which approaches with increasing $N$ a power law exponent $-\Theta_2=-3/4$ 
(dashed line) corresponding to a compact object of fractal line dimension $\dpm =d - \Theta_2 = 5/4$.
Also included is the Porod scattering expected for a compact 2d object with smooth perimeter
(dash-dotted line) and the Fourier transform of the Redner-des Cloizeaux approximation 
(solid line). 
The increase of the scattering for large $q$ (Bragg peak) is due to the packing of the beads
on local scale. Only in this limit does $F(q)$ become chain length independent.
\label{fig_Fq}
}
\end{figure}

{\em Intrachain form factor.}
Neither segmental size distribution nor bond-bond correlation function are readily accessible experimentally.
It is thus important that $\Theta_2$ should be measurable --- at least in principle --- from an analysis of 
the intrachain form factor
$F(q) = \frac{1}{N} \sum_{n,m=1}^{N} \langle \exp\left[i \vec{q} \cdot (\vec{r}_n-\vec{r}_m) \right]\rangle$.
The reason for this is that the form factor can be expressed by the
Fourier transform 
$G(q,s) =  
\langle \exp\left[i \vec{q} \cdot (\vec{r}_n-\vec{r}_m) \right]\rangle_{nm}$
of the segmental size distribution $G(r,s)$ with $\vec{q}$ being the wavevector:
$F(q) = \frac{2}{N} \int_0^N ds (N-s) G(q,s)$. 
Assuming $G(r,s) \approx G_2(r,N)$ and using the Redner-des Cloizeaux approximation
(Eq.~(\ref{eq_theta})) this yields a lengthy analytic
formula (not given) which is represented by the solid line in Fig.~\ref{fig_Fq} \cite{foot_approx}. 
For wavevectors corresponding to the power-law regime of Eq.~(\ref{eq_theta}) 
this reduces to the simple power law
\begin{equation}
F(q)/N  \approx 1.98/(q \Rgyr(N))^{d+\Theta_2}
\label{eq_Fq}
\end{equation}
indicated by the dashed line.
Note that the above scaling is a direct consequence of $G(r,s) \sim r^{\Theta_2}$ and does not rely on the 
Redner-des Cloizeaux approximation.
We rescale the wavevector with the measured radius of gyration $\Rgyr(N)$ (presented in Fig.~\ref{fig_sum}) 
to collapse all data in the Guinier regime for small $x=q\Rgyr(N) \ll 1$ and use a Kratky representation for 
the vertical axis $y=(F(q)/N) x^2$. 
While $y$ becomes constant and independent of chain length for $x \gg 1$ for Gaussian chains 
(as shown by the Debye formula indicated) \cite{DegennesBook},
we observe over a decade in $x$ a striking non-monotonous behavior.
Our data suggests that Eq.~(\ref{eq_Fq}) is approached systematically with increasing chain length ---
the central numerical result presented in this paper.
%

%
%%%%%%%%%%%%%%%%%%%%%%%%%%%%%%%%%%%%%%%%%%%%%%%%%%%%%%%%%%%%%%%%%%%%%%%%%%%%%%%%%%%%%%%%%%%%%%%%%%%%%%%%%%%%%%%
%
{\em Identification of $\Theta_2$ and the fractal line dimension.}
The preceding discussion focused exclusively on intrachain properties. 
Since 2d chains are compact (Fig.~\ref{fig_sum}) only monomers on the chain perimeter interacting with monomers
from other chains can contribute to the scattering. Quite generally, the scattering intensity $N F(q)$ 
of compact objects becomes proportional to the mean ``surface" $L(N) \sim R(N)^{\dpm}$
for $q R(N) \ll 1$ %with $R(N)$ being the typical size. This implies 
which implies
the generalized Porod law \cite{BenoitBook,bale84,bray88} 
\begin{equation}
N F(q) \approx N^2 / \left(q R(N) \right)^{2d-\dpm}.
\label{eq_Fqsurface}
\end{equation}
For a 2d object with smooth perimeter ($\dpm=1$) this corresponds to the classical Porod scattering 
$F(q) \sim 1/q^3$ represented by the dash-dotted line in Fig.~\ref{fig_Fq}.
Comparing Eq.~(\ref{eq_Fqsurface}) with Eq.~(\ref{eq_Fq}) shows that 2d melts are characterized
by a fractal line dimension $\dpm = d - \Theta_2$ and demonstrates finally the scaling of the 
perimeter length $L(N)$ postulated in the Introduction and verified numerically in Fig.~\ref{fig_sum}. 
By labeling only the monomers of sub-chains 
(which corresponds to a scattering amplitude $s F(q) \sim L(s) \sim R(s)^{\dpm}$)
the above argument is readily generalized to the perimeter 
length $L(s)$ of arbitrary segment length $s \le N$ \cite{foot_Lsalternative}.
%

%%%%%%%%%%%%%%%%%%%%%%%%%%%%%%%%%%%%%%%%%%%%%%%%%%%%%%%%%%%%%%%%%%%%%%%%%%%%%%%%%%%%%%%%%%%%%%%%%%%%%%%%%%%%%%
{\em Summary.}
Investigating various static properties of linear polymer melts in 2d we have demonstrated that 
the compact chains and chain segments ($\nu=1/2$) are characterized by a fractal perimeter $L(s)$ 
of line dimension $\dpm=d-\Theta_2=5/4$. 
As may be seen from Fig.~\ref{fig_Fq}, computationally very demanding systems with chain length
$N > 10^3$ are required, and have thus been simulated, to put to the test the suggested scaling behavior.
Our results may be verified experimentally from the scaling of the intrachain form factor $F(q)$ 
whose Kratky representation is predicted to reveal a strong non-monotonous behavior.
This should also hold in semidilute solutions provided that the chains are long enough
\cite{foot_para}.
Interestingly, the fractality of the perimeter precludes a finite line tension.
Thus, shape 
fluctuations of the segments are not suppressed exponentially \cite{carm}, but may occur in an 
``amoeba-like'' fashion by advancing and retracting ``lobes''.
According to a recent suggestion \cite{semenov} the 
relaxation time $\tau(s)$ of a chain segment may, hence, scale as 
$\tau(s) \sim L(s)^3 \sim s^{15/8}$ rather than as $s^2$ as in the standard Rouse model
\cite{DoiEdwardsBook}.
Clearly, as Gaussian chain statistics is inappropriate to describe conformational properties of
2d melts, there is no reason why a model based on this statistics should allow to describe, 
e.g., the perimeter length fluctuations. 
Since the latter property is in principle accessible experimentally
from the dynamical intrachain structure factor
$F(q,t)$ \cite{BenoitBook,DoiEdwardsBook}
this is an important issue we are currently investigating.

%
%%%%%%%%%%%%%%%%%%%%%%%%%%%%%%%%%%%%%%%%%%%%%%%%%%%%%%%%%%%%%%%%%%%%%%%%%%%%%%%%%%%%%%%%%%%%%%%%%%%%%%%%%%%%%%
{\em Acknowledgements.}
We thank  the ULP, the IUF, the Deutsche Forschungsgemeinschaft (KR 2854/1-1) and the ESF-STIPOMAT programme
for financial support. A generous grant of computer time by the IDRIS (Orsay) is also gratefully acknowledged.
We are indebted to A.N.~Semenov for helpful discussions.

%%%\bibliography{../bibl,foot,../../bibl/BOOK,../../bibl/JPW,../../bibl/ANS}

\begin{thebibliography}{23}
\bibitem[{foo({\natexlab{c}})}]{foot_Ldistance}
\bibinfo{note}{We have checked that the scaling of $L(s)$ does not depend on
  the distance used for defining a perimeter monomer.}
\bibitem[{foo({\natexlab{d}})}]{foot_approx}
\bibinfo{note}{Since we use only the $\Theta_2$-exponent and in addition to
  this the Redner-des Cloizeaux formula there are two levels of approximation.
  Our final result is rephrased consistently in terms of $x=q\Rgyr(N)$ and,
  thus, by construction our formula yields correctly both the Guiner regime for
  $x\ll 1$ and the generalized Porod limit for $x \gg 1$, but not necessarily
  the hump region at $x \approx 2$. In practice, our approximation is fine for
  all $x$ as shown by Fig.~\ref{fig_Fq}.}
\bibitem[{foo({\natexlab{e}})}]{foot_Lsalternative}
\bibinfo{note}{An alternative derivation of Eq.~(\ref{eq_fractal}) is obtained
  from a scaling argument using the return probability $1/s^{1+\nu\Theta_2}$
  measured in Fig.~\ref{fig_eiej}. The key point is that a monomer in a long
  segment cannot ``distinguish'' if the contact is realized through the
  backfolding of its own segment or by another segment \cite{dupl,semenov}.
  Since the probability of such a binary contact, $L(s)/s$, must be
  proportional to the return probability times the number $s$ of monomers in
  the second segment this yields Eq.~(\ref{eq_fractal}).}
\end{thebibliography}

\end{document}